\documentclass[12pt, a4paper]{article}
\usepackage[utf8]{inputenc}

\usepackage{amsmath,amssymb,graphicx}

\usepackage[margin=2.5cm]{geometry}
\usepackage{enumitem}
\usepackage[round]{natbib}
\usepackage{setspace}
\usepackage{hyperref}
\usepackage{doi}

\renewcommand*{\theta}{\vec\vartheta}

\title{Approaches Toward the Bayesian Estimation of the Stochastic Volatility Model with Leverage}
\author{Darjus Hosszejni\footnote{WU Vienna University of Economics and Business, Austria. Email: \href{mailto:darjus.hosszejni@wu.ac.at}{darjus.hosszejni@wu.ac.at}.} \and Gregor Kastner\footnote{WU Vienna University of Economics and Business, Austria. Email: \href{mailto:gregor.kastner@wu.ac.at}{gregor.kastner@wu.ac.at}.}}
\date{February 1, 2019}
\begin{document}

\maketitle
\begin{center}
\textbf{With very minor editorial changes, this article is published as:}\\
Darjus Hosszejni and Gregor Kastner. Approaches toward the Bayesian estimation of the stochastic volatility model with leverage. In Raffaele Argiento, Daniele Durante, and Sara Wade, editors, \textit{Bayesian Statistics and New Generations -- Selected Contributions from BAYSM 2018}, volume 296 of \textit{Springer Proceedings in Mathematics \& Statistics}, pages 75--83, Cham, 2019. Springer. \href{https://doi.org/10.1007/978-3-030-30611-3_8}{doi:10.1007/978-3-030-30611-3\_8}.
\end{center}

\vspace{3em}

\begin{abstract} \noindent
The sampling efficiency of MCMC methods in Bayesian inference for stochastic volatility (SV) models is known to highly depend on the actual parameter values, and the effectiveness of samplers based on different parameterizations varies significantly. We derive novel algorithms for the centered and the non-centered parameterizations of the practically highly relevant SV model with leverage, where the return process and innovations of the volatility process are allowed to correlate. Moreover, based on the idea of ancillarity-sufficiency interweaving (ASIS), we combine the resulting samplers in order to guarantee stable sampling efficiency irrespective of the baseline parameterization.We carry out an extensive comparison to already existing sampling methods for this model using simulated as well as real world data.
\end{abstract}


\section{Introduction and Model Specification} \label{sec:intro}
Stochastic volatility (SV) models \citep{taylor1982sv} are an increasingly popular choice for modeling financial return data.
The basic SV model assumes an autoregressive structure for the log-volatility, and it is able to match the empirically observable low serial autocorrelation in the return series but high serial autocorrelation in the squared return series.
The SV model with leverage \citep[SVL, ][]{harvey1996estimation} extends the SV model by allowing the return series and the increment of the log-volatility series to correlate.
This correlation models a real world phenomenon, the asymmetric relationship between returns and their volatility.

SVL, in its centered parameterization (C), is typically formulated as
\begin{equation}\label{eqn:c}
    \begin{split}
        y_t &= \exp(h_t/2)\varepsilon_t, \\
        h_{t+1} &= \mu+\varphi(h_t-\mu)+\sigma\eta_t, \\
        \text{cor}(\varepsilon_t,\eta_t) &= \rho,
    \end{split}
\end{equation}
for $t=1,\dots,T$, where $\varepsilon_t,\eta_t\sim\text{ i.i.d. }\mathcal{N}(0,1)$.
The only observed variable is $\vec y=(y_1,\dots,y_T)^\prime$, usually some de-meaned financial return series.
An AR(1) structure is assumed for the latent log variance $\vec h=(h_1,\dots,h_T)^\prime$, with mean $\mu$, persistence $\varphi$, and increment volatility $\sigma$.
The leverage effect is captured by $\rho$, which is zero in the basic SV model by assumption.
\newcommand*{\pC}{p_\text{C}}

An equivalent specification, called the non-centered parameterization (NC), can be obtained by substituting $\tilde h_t=(h_t-\mu)/\sigma$ into~(\ref{eqn:c}), thereby moving $\mu$ and $\sigma$ from the state equation to the observation equation. The resulting formulation is given by
\begin{equation}\label{eqn:nc}
    \begin{split}
        y_t &= \exp((\mu+\sigma\tilde h_t)/2)\varepsilon_t, \\
        \tilde h_{t+1} &= \varphi\tilde h_t+\eta_t.
    \end{split}
\end{equation}

Common priors are chosen from the literature: $(\varphi+1)/2 \sim \text{Beta}(a_\varphi,b_\varphi)$, $(\rho+1)/2 \sim \text{Beta}(a_\rho,b_\rho)$, $\sigma^2 \sim \text{Gamma}(\alpha_\sigma,\beta_\sigma)$, $\mu \sim \mathcal{N}(\mu_\mu,\sigma_\mu^2)$, $h_1 \sim \mathcal{N}(\mu,\sigma^2/(1-\varphi^2))$ \citep{fruhwirth2010stochastic,Kastner2014,Omori2007}.

While the SV model is accessible through the R \citep{rlanguage} package \texttt{stochvol} \citep{kastner2016dealing}, it does not cater for the leverage effect, and, to the best of our knowledge, there is no implementation of SVL that works out-of-the-box in a free, open source environment.
Our goal is to extend the package with an easy-to-use MCMC sampler that performs reasonably well on a diverse variety of data sets.
To this end, we compare various sampling algorithms through a large simulation study from a practical viewpoint.
In doing so, it is important to note that \texttt{stochvol} is often used as a subsampler for hierarchical models such as (vector auto) regressions or multivariate (factor) SV models. Consequently, in order to use the extended package in a similar manner, adaptive algorithms are not preferred, as their adaptation state can be cumbersome to implement within a larger MCMC scheme.

\section{Estimation Strategies} \label{sec:est}
The state-of-the-art solution \citep{Omori2007} for estimating $\vec{h}$ is based on linearizing the observation equation in~\eqref{eqn:nc}, and employing a ten-component bivariate Gaussian mixture approximation to the joint law of $(\log\varepsilon_t^2,\eta_t)$ separately for each time point, thus introducing a new array of latent variables $s_t\in\{1,\dots,10\}$, $t=1,\dots,T$, encoding the mixture components.
The resulting conditionally Gaussian state space can be written as
\begin{equation}\label{eqn:aux}
    \begin{split}
        y_t^\ast &= \mu+\sigma\tilde h_t+m^{(1)}_{s_t}+v^{(1)}_{s_t}w_t, \\
        \tilde h_{t+1} &= \varphi\tilde h_t+\sqrt{1-\rho^2}z_t + d_t\rho\left(m^{(2)}_{s_t}+v^{(2)}_{s_t}w_t\right),
    \end{split}
\end{equation}
where $y_t^\ast=\log(y_t^2)$, $d_t=\text{sgn}(y_t)$, $w_t$, $z_t\sim\text{ i.i.d. }\mathcal{N}(0,1)$ for $t=1,\dots,T$, and $m^{(i)}_{j}$, $v^{(i)}_{j}$ are model-independent constants for $i=1,2$, and $j=1,\dots,10$, defined in \citet{Omori2007}.
\newcommand*{\pA}{p_\text{A}}

Let $\theta=(\varphi,\rho,\sigma,\mu)^\prime$, and $\vec s=(s_1,\dots,s_T)^\prime$. The sampling algorithm of the auxiliary model (AUX) consists of repeating the steps below.
\begin{enumerate}[label=\textbullet,ref=\arabic*]
  \item\label{aux:s1} Step 1: Draw $\vec s\mid\vec y,\vec h,\theta$ using inverse transform sampling with the posterior probabilities calculated as in Section 2.3.2 of \citet{Omori2007}.
  \item\label{aux:s2} Step 2: Draw $\varphi,\rho,\sigma\mid\vec y,\vec s$ via an independent Metropolis-Hastings (MH) step utilizing the Laplace approximation of the collapsed distribution of $\varphi,\rho,\sigma\mid\vec y,\vec s$ as the proposal. The calculation of the acceptance ratio includes Kalman filter evaluations, numerical optimization, and numerical differentiation.
  \item\label{aux:s3} Step 3: Draw $\mu\mid\vec y,\vec s,\varphi,\rho,\sigma$, and then $\vec h\mid\vec y,\vec s,\theta$, using Gaussian simulation smoothing \citep{carter1994gibbs,fruhwirth1994data}.
\end{enumerate}
At least three issues arise with this method.
First, due to the involvement of Kalman filter evaluations and the numerical optimization part in Step~\ref{aux:s2}, the execution time of the sampler is significantly worse than the runtime of methods with more na\"ive proposals, e.g.\ MH algorithms based on sampling from the full conditional distribution. According to our measurements, Step~\ref{aux:s2} requires around 80\% of the total runtime.
Second, for extreme data sets, the sampler might get stuck in a state and be unable to accept a new state for many iterations.
Third, and finally, the numerical optimization step is sensitive to its configuration, possibly returning a negative semi-definite Hessian matrix at the found mode.

Hence, for parameter sampling, we replace Step~\ref{aux:s2} by a random-walk MH (RWMH) method which estimates~\eqref{eqn:c} or~\eqref{eqn:nc} without resorting to the auxiliary mixture approximation.
For the latent vector, we again use the highly efficient Step~\ref{aux:s3} of AUX as a proposal, followed by an MH acceptance-rejection step to correct for the difference between models~\eqref{eqn:c} and~\eqref{eqn:aux}.

As already shown for SV \citep{Kastner2014,McCausland2011}, samplers based on different parameterizations can have substantially different sampling efficiencies on the same data set due to the altered dependence structure.
To exploit this phenomenon, the ancillarity-sufficiency interweaving strategy \citep[ASIS, ][]{Yu2011} can utilize samplers of both C and NC, and thus ASIS may be able to deliver a markedly higher effective sample size than samplers based on a single parameterization only.

The RWMH sampling algorithm estimates SVL by repeating the steps below.
\begin{enumerate}[label=\textbullet,ref=\arabic*]
  \item\label{rwmh:s1} Step 1: Draw $\vec h\mid\vec y,\theta$.
    A candidate $\vec h^\ast$ is proposed using the AUX sampler by drawing $\vec s\mid\vec y,\vec h,\theta$ and then drawing $\vec h\mid\vec y,\vec s,\theta$ as explained in Steps~\ref{aux:s1} and~\ref{aux:s3} of algorithm AUX.
    Subsequently, $\vec h^\ast$ is accepted with probability
    \begin{equation*}\label{eqn:acc}
      \min\left\{1,\frac{\pC\left(\vec h^\ast\mid\vec y,\theta\right)}{\pC\left(\vec h\mid\vec y,\theta\right)} \frac{\pA\left(\vec h\mid\vec y,\theta\right)}{\pA\left(\vec h^\ast\mid\vec y,\theta\right)}\right\},
    \end{equation*}
    where $p_C$ and $p_A$ denote the corresponding posteriors resulting from specifications \eqref{eqn:c} and \eqref{eqn:aux}, respectively.
  \item\label{rwmh:s2} Step 2: Draw $\theta\mid\vec y,\vec h$.
    In order to avoid a possibly problematic truncation of the proposal distribution, the parameter vector $\theta$ is transformed from $(-1,1)\times (-1,1)\times (0,\infty)\times\mathbb{R}$ to $\mathbb{R}^4$ by applying the transformation $x\mapsto 0.5\log((1+x)/(1-x))$ to $\varphi$ and $\rho$, and by taking the logarithm of $\sigma^2$.
    Then, in the resulting unbounded space, a simple four-dimensional Gaussian random walk is proposed. Its innovation covariance matrix elements are fixed at $0.1$ on the diagonal and zero elsewhere.
  \item\label{rwmh:s3} If ASIS is applied, then, after Step~\ref{rwmh:s2}, $\tilde{\vec h}=(\tilde h_1,\dots,\tilde h_T)^\prime$ is calculated using the new values of $\sigma$, $\mu$, followed by a new draw from $\theta\mid\vec y,\tilde{\vec h}$.
    Finally, in order to move back to the original parameterization, $\vec h$ is recalculated from $\tilde{\vec h}$ and the new values of $\sigma$, $\mu$.
\end{enumerate}

ASIS is a natural extension to the RWMH samplers for the centered and the non-centered parameterizations.
However, in the case of AUX, resampling in a different parameterization is detrimental to sampling efficiency for two reasons.
First, in Step~\ref{aux:s2}, the parameters $\varphi$, $\rho$ and $\sigma$ are drawn from a collapsed distribution that is independent of $\vec h$. Consequently, ASIS provides only negligible gains in sampling efficiency.
Second, if ASIS were applied to AUX, the computationally most expensive parts of Step~\ref{aux:s2} would be repeated, thereby increasing the execution time by around 80\%.

\section{Simulation Study} \label{sec:simres}
In order to assess the efficiency of our estimation algorithms for the parameter vector $\theta$, we simulate data using SVL from an extensive grid of data generating processes (DGPs).
The parameters $\varphi_\text{true}$, $\rho_\text{true}$, $\sigma_\text{true}$ vary on a $\{0,0.5,0.9,0.95,0.99\}\times\{-0.6,-0.3,0,0.3,0.6\}\times\{0.1,0.3,0.5\}$ grid.
For the sake of readability, $\mu_\text{true}$ is set to $-9$ in all cases, resulting in $75$ distinct parameter settings. This choice covers previously investigated ranges \citep{Jacquier2004,Kastner2014}.
After the burn-in, respectively, adaptation phase, $50\,000$ MCMC draws are obtained from the posterior distribution.
We repeat this exercise for ten data sets of length 300, and ten data sets of length 3000, for eight sampling algorithms: AUX, Stan-C, Stan-N, JAGS-C, JAGS-N, RWMH-C, RWMH-N, and RWMH-ASISx5, where C and N stand for the centered and, respectively, non-centered parameterization, while ASISx5 denotes the algorithm repeating the two steps of ASIS five times after each draw of $\vec h$, which in general we found to be superior to executing the two ASIS steps only once.
Note, that, although they do not fit our needs due to their adaptation phase, we include Stan \citep{stan} and JAGS \citep{jags} as benchmarks, and all reported results are based on the chain after adaptation has stopped.
We fix the priors throughout the simulation study to $a_\varphi=20$, $b_\varphi=1.5$, $a_\rho=3$, $b_\rho=6$, $\alpha_\sigma=0.5$, $\beta_\sigma=0.5$, $\mu_\mu=-10$, and $\sigma^2_\mu=100$.
The prior hyperparameters of $\varphi$, $\sigma^2$, and $\mu$ are chosen from previous studies \citep{Kastner2014,Nakajima2009}, and the slightly informative prior on $\rho$ is chosen to improve the estimation process of Stan-C and of AUX in the extremes of the parameter grid.
However, results not reported here due to space constraints indicate that when $T=300$, the posterior distribution of $\rho$ is only mildly affected by this choice compared to a uniform prior, whereas when $T=3000$ the differences are barely noticeable.

The resulting $12\,000$ MCMC chains were computed on a cluster of computers consisting of 400 Intel E5 2.3 GHz cores running R version 3.4.3.
The Stan and the JAGS models were estimated using \texttt{rstan} \citep{stan} version 2.17.3 and \texttt{rjags} \citep{jags} version 4-6.
The RWMH samplers and AUX were based on our computationally efficient \texttt{Rcpp} \citep{rcpp} implementation.
The typical runtime of the samplers is summarized in Table~\ref{tab:run}.
Inefficiency factors and effective sample sizes were calculated using the \texttt{coda} \citep{coda} package, data analysis and visualization was done with the help of the \texttt{tidyverse} \citep{rtidyverse} package.

\begin{table}[t]
  \centering
  \begin{tabular}{ccccccc}
    Stan-C       & Stan-N & JAGS-C & JAGS-N & RWMH  & RWMH-ASISx5 & AUX   \\ \hline
    90--642 & 59--441 & 22--31  & 50--106  & 6--21 & 14--29       & 44--86
  \end{tabular}
  \caption{Typical execution times (in minutes) for $50\,000$ draws after the burn-in when $T=3000$.
The displayed values correspond to the first and the ninth deciles of all wall clock times.
The choice of the parameterization affects the execution time when JAGS or Stan is used and thus in these cases runtimes are shown separately for C and NC.}
  \label{tab:run}
\end{table}

We assess the statistical efficiency of the different competitors through their inefficiency factor (IF), an estimator for the integrated autocorrelation time $\tau$, given by $\tau=1+2\sum_{t=1}^\infty\rho_\text{auto}(t)$, where $\rho_\text{auto}(t)$ denotes the autocorrelation function at lag $t$.
For an MCMC sample $S$, the IFs reported here are calculated as $\text{IF}_S=n_S/\text{ESS}_S$ \citep{Kastner2014}, where $n_S$ is the size of $S$, and $\text{ESS}_S$ stands for the effective sample size of $S$, the size of a serially uncorrelated sample having the same Monte Carlo standard error as $S$.
A good sampler has low serial correlation, thus the aim is to provide samples with low IF, or, in other words, high ESS.
In practice, computational speed is comparably important to computational efficiency.
Hence, the final assessment is based on the effective sampling rate (ESR), defined as the ESS divided by the execution time.
We note that incorporating runtime in the assessment of algorithms may be problematic due to inconsistent implementations \citep{kriegel2016runtime}; however, as one of our objectives is a software package, we consider the computational speed an essential part of our study.

\subsection{Collapsed vs.\ Full Conditional Sampling}
\begin{figure}[t]
  \includegraphics[width=\textwidth]{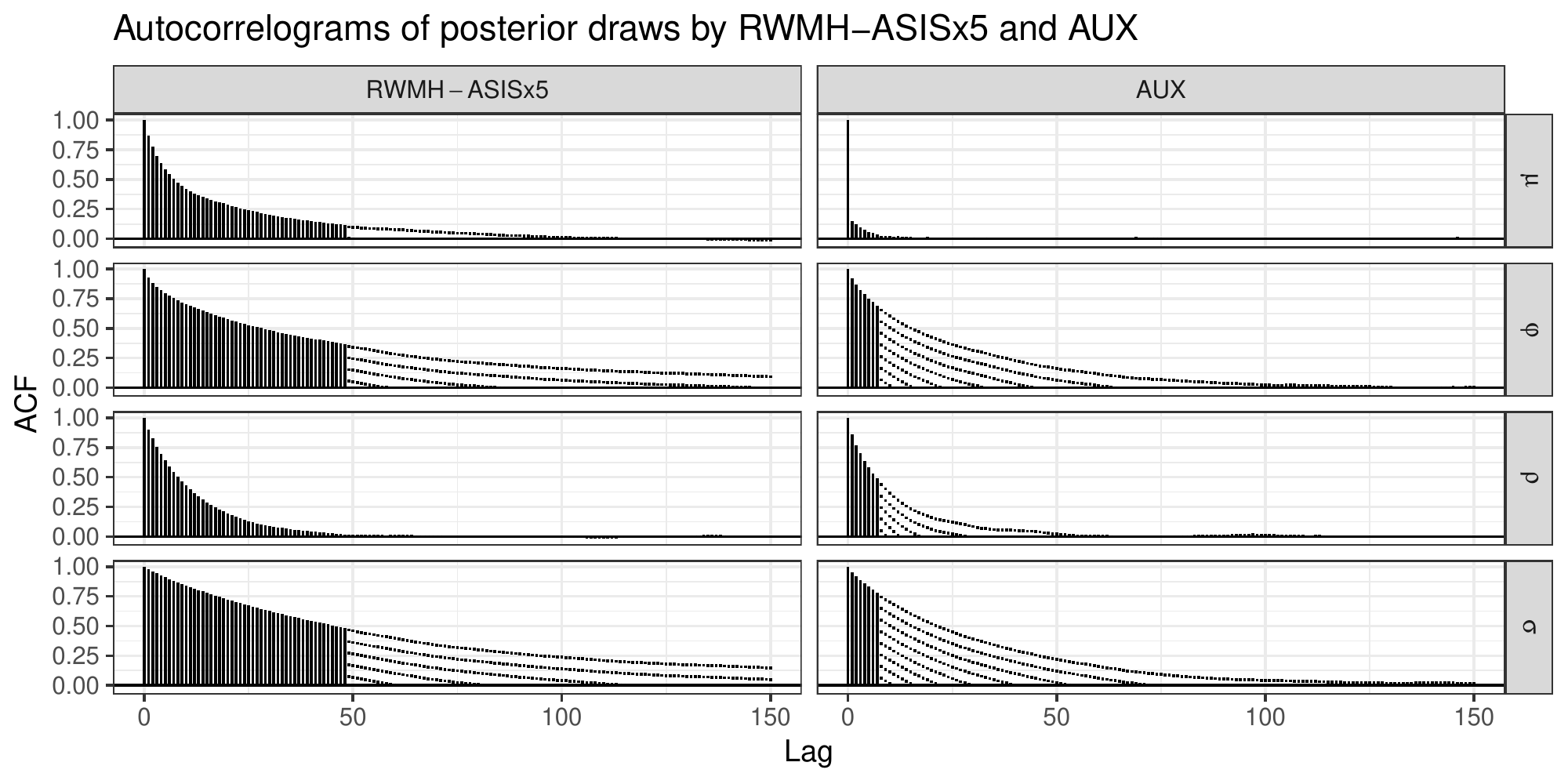}
  \caption{Autocorrelation functions of the posterior draws for $\mu$, $\varphi$, $\rho$, and $\sigma$, using RWMH-ASISx5 and AUX, for an illustrative example where $\varphi_\text{true}=0.95$, $\rho_\text{true}=-0.3$, $\sigma_\text{true}=0.3$, and $T=300$.
  The line type indicates the speed of the Monte Carlo simulation: the number of solid lines equals the average number of samples drawn in $0.1$ seconds.}
  \label{fig:acf}
\end{figure}

AUX employs a well-known technique for improving the statistical efficiency of MCMC simulations by using a collapsed distribution for sampling $\varphi$, $\rho$, and $\sigma$.
This means that some variables are marginalized out in order to decrease the serial dependence in the chain \citep{Liu1994}.
ASIS, on the other hand, takes advantage of being able to reorganize the dependence.
Which technique is superior in practice largely depends on computational aspects.
Figure~\ref{fig:acf} exemplifies the problem by displaying the autocorrelograms of the outputs of RWMH-ASISx5 and AUX, for the parameters $\mu$, $\varphi$, $\rho$, and $\sigma$, based on a selected data set.
The figure illustrates the execution time as well: in both columns, the number of solid lines indicates the average number of samples drawn in $0.1$ seconds.
Thus, in each facet, the height of the rightmost solid line visualizes the ESR for the given parameter and sampler.
Although the autocorrelation functions of AUX decay faster than the ones of RWMH-ASISx5, the latter counterbalances its disadvantages by its speed.
Note, however, that different DGPs tend to produce qualitatively different pictures, making the choice between AUX and RWMH-ASISx5 non-trivial.

\subsection{Efficiency Overview}
\begin{figure}[t]
  \includegraphics[width=\textwidth]{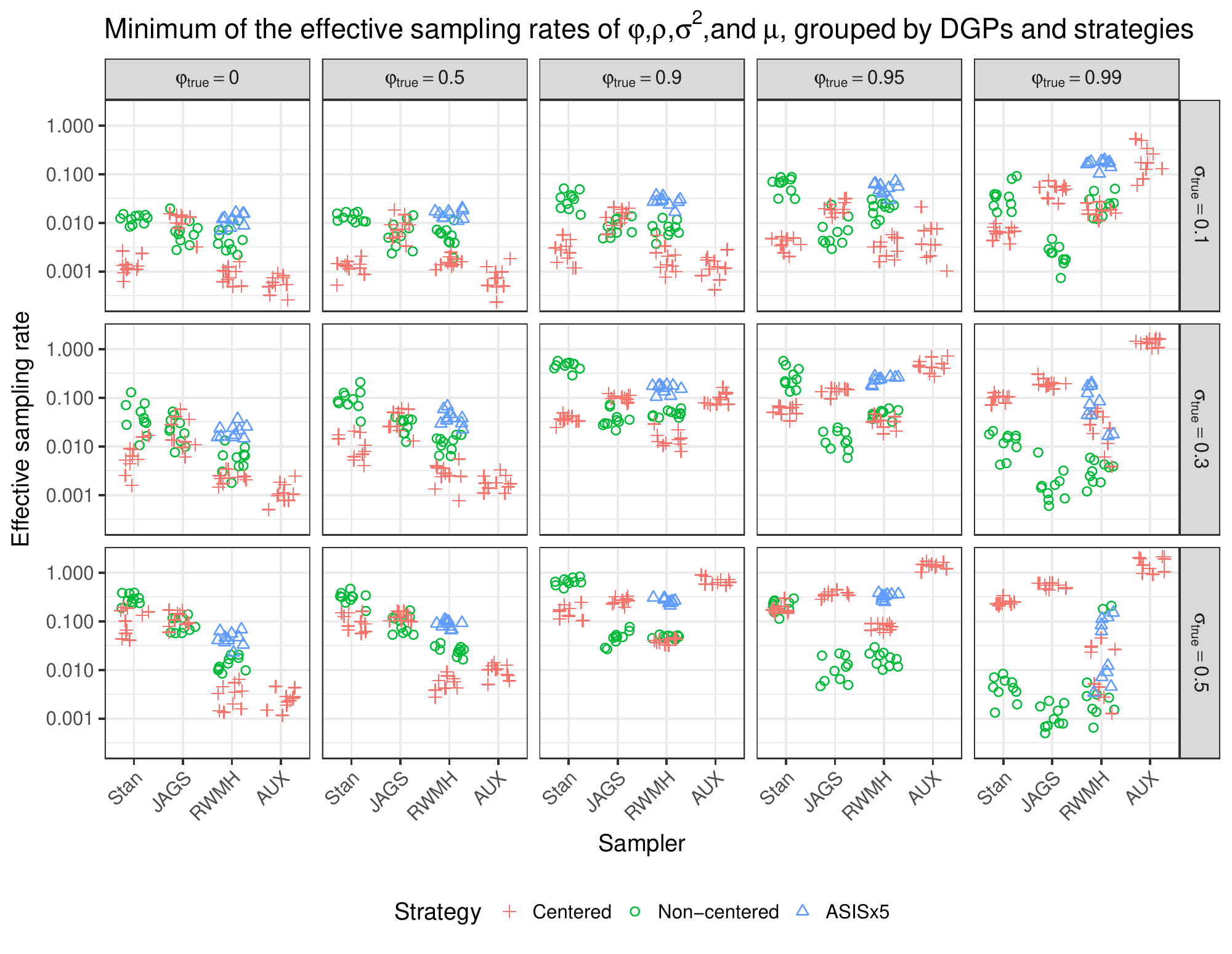}
  \caption{Minimal effective sampling rates of all the examined samplers and strategies, for the whole range of $\varphi_\text{true}$ and $\sigma_\text{true}$ values, while, for the sake of readability, $\rho_{true}$ is set to $-0.3$, and $T$ to $3000$.
  In each facet, there are 10 data points plotted for each sampler and strategy, corresponding to the 10 repetitions of the simulation exercise.
  A small horizontal noise has been applied to the position of the points to increase their visibility.}
  \label{fig:esr}
\end{figure}

The minimal ESR is the minimum taken over the ESRs of $\varphi$, $\rho$, $\sigma$, and $\mu$, and, thus, it measures the speed of discovering the joint posterior $p(\theta\mid\vec y)$.
In order to provide an overview, Figure~\ref{fig:esr} displays the minimal ESRs for each sampler and strategy, and for all DGPs with $\rho_\text{true}=-0.3$ and $T=3000$.
Taking into account that Stan and JAGS are general-purpose probabilistic modeling frameworks, they perform surprisingly well compared to AUX and our RWMH implementations which have been developed specifically for the model at hand.
However, the absence of a best performing method is eye-catching.
In particular, the choice between AUX and RWMH is noticeably difficult.

In terms of variability, note that the ESRs of AUX range from below $0.001$ to above $1$, while the ESRs of RWMH-ASISx5 fall between $0.01$ and $0.1$. This renders the latter more stable by around two orders of magnitude.

\section{Application to Financial Data Sets} \label{sec:realres}
We apply the eight estimation methods to seven univariate time series of daily financial log-returns covering four asset types and two economic periods.
The first time interval is a booming, pre-crisis period starting from 2005-01-01 and ending on 2007-12-31, including a total of 872 business days.
The second interval is a more recent, more volatile period between 2015-01-01 and 2018-06-30, including 1014 business days.
The series under consideration are the Bitcoin price in USD (ticker: BTCUSD=X), hereafter BTC, the German DAX index (ticker: \^{}GDAXI), the EUR/USD exchange rate (ticker: EURUSD=X), hereafter EUR, and a large German company, the Merck KG's equity (ticker: MRK.DE), hereafter MRK.
The data is provided by Yahoo! Finance.

\begin{figure}[t]
  \includegraphics[width=\textwidth]{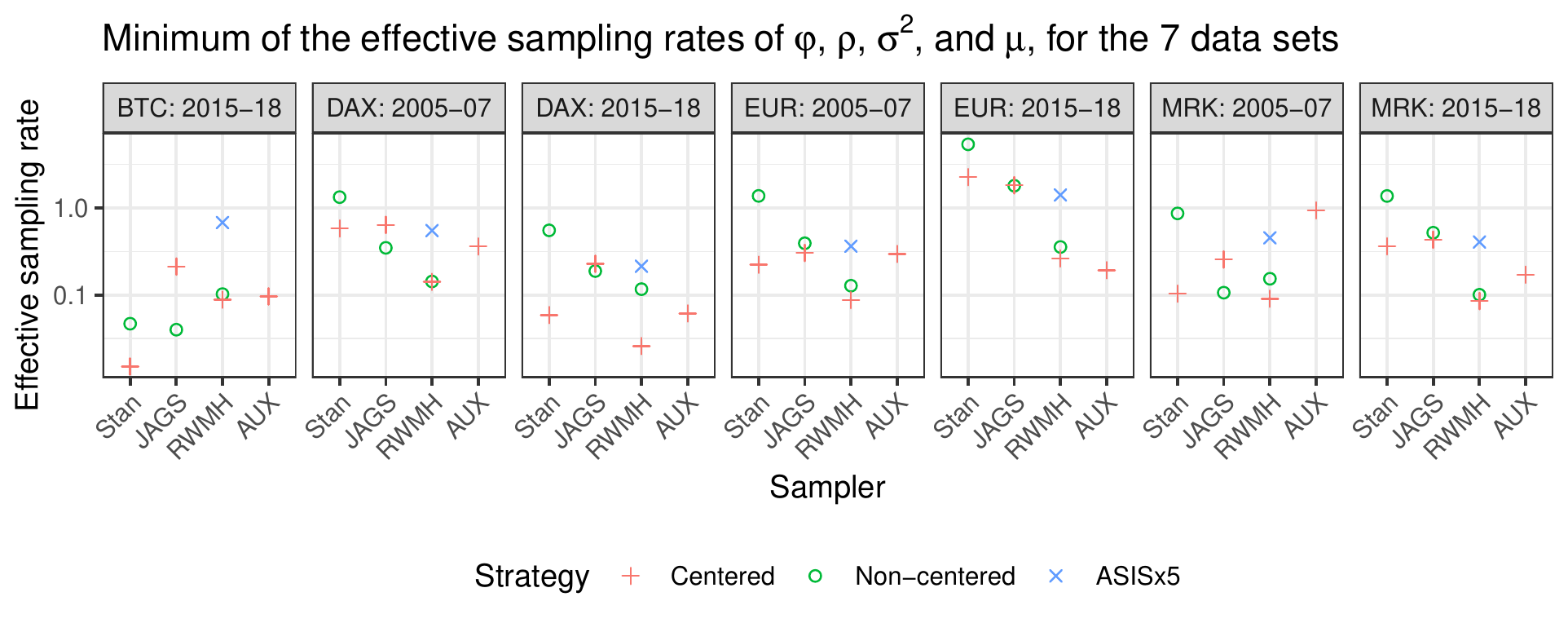}
  \caption{Minimal effective sampling rate for seven data sets.
  Facets correspond to time series, where ``2005--07'' and ``2015--18'' denote the first and the second time period, respectively.
  In each facet, each point corresponds to a certain sampler and strategy.}
  \label{fig:app}
\end{figure}

Figure~\ref{fig:app} summarizes the results of the exercise, carried out under the same prior specification as in Section~\ref{sec:simres}.
It is interesting to note that Stan generally shows high ESRs with the only exception of BTC where RWMH-ASISx5 excels.
Focusing on the comparison of RWMH and AUX it stands out that without interweaving, AUX is generally to be preferred, whereas RWMH-ASISx5 tends to outperform AUX in all scenarios but one.
The overall picture is similar to Figure~\ref{fig:esr}, as there is no single algorithm that dominates on all data sets.

\section{Conclusion}
The paper at hand contributes to the literature on MCMC sampling algorithms by investigating the efficiency of several competing methods for the stochastic volatility model with leverage.
We derived an RWMH sampler and improved it through ASIS and an efficient latent state sampler.
Moreover, we carried out a computational experiment to compare our novel method to the state-of-the-art approach, an auxiliary mixture sampler, and to Stan and JAGS implementations as benchmarks.
Based on our results, we conclude that employing our boosted na\"ive estimator for the latent space stabilizes the effective sampling rate of the algorithm by avoiding numerical optimization and differentiation.

Current research is directed towards further financial applications including factor models \citep{kastner2017efficient}, and extending the R package \texttt{stochvol} to allow for leverage.

\bibliography{references}

\end{document}